\documentclass[10pt]{iopart}

\begin{document}

\def\Dirac{i\partial\!\!\!\!/}
\def\DDirac{iD\!\!\!\!/}
\def\dirac{i\partial\!\!\!\!/-eA\!\!\!\!/}

\newcommand{\gc}{\gamma _M^0}
\newcommand{\gu}{\gamma _M^1}
\newcommand{\gd}{\gamma _M^2}
\newcommand{\gce}{\gamma _0}
\newcommand{\gue}{\gamma _1}
\newcommand{\beq}{\begin{eqnarray}}
\newcommand{\eeq}{\end{eqnarray}}
\newcommand{\nn}{\nonumber}

\makeatletter
 \renewcommand{\theenumi}{\arabic{enumi}}
\article[Relativistic quantum Hall effect]{}{Finite-temperature relativistic Landau problem and the relativistic quantum Hall effect}
\author{C G Beneventano\footnote{Member of CONICET} and
E M Santangelo\footnote{Member of CONICET}}
\address{Departamento de F{\'\i}sica, Universidad Nacional de La Plata
\\Instituto de F{\'\i}sica de La Plata, UNLP-CONICET\\
C.C. 67, 1900 La Plata, Argentina}
\ead{\mailto{gabriela@obelix.fisica.unlp.edu.ar},
\mailto{mariel@obelix.fisica.unlp.edu.ar}}
\begin{abstract}
This paper presents a study of the free energy and particle density of the relativistic Landau problem, and their relevance to the quantum Hall effect. We study first the zero temperature Casimir energy and fermion number for Dirac fields in a 2+1-dimensional Minkowski space-time,in the
presence of a uniform magnetic field perpendicular to the spatial
manifold. Then, we go to the finite-temperature problem, with a
chemical potential, introduced as a uniform zero component of
the gauge potential. By performing a Lorentz boost, we obtain Hall's conductivity in the case of crossed electric and magnetic fields. \\[.3cm]
{\bf Subject Classification} 

\end{abstract}

\section{Introduction}\label{intr}

The quantization of the Hall conductivity \cite{klitzing} is a remarkable quantum phenomenon,
which occurs in two-dimensional electron systems, at low temperatures and strong perpendicular
magnetic fields. Most proposed explanations for this phenomenon (for a review see, for instance, \cite{basics}) rely on one-particle theory and make use of the Kubo formula concerning the conductivity as a linear response function to the external field \cite{ando}.

It is the aim of this paper to show that, in the context of relativistic field theory, the quantization of the Hall conductivity in multiples of the magnetic flux arises as a consequence of the spin-statistics theorem, which is a straightforward outcome of such a theory.

After this calculation was finished, a very interesting, recently published \cite{gusynhall}, calculation of the Hall conductivity in relativistic systems was brought to our knowledge. We will compare our results to those obtained in this recent publication in our final comments.

In section \ref{Sect2} we present the theory of Dirac fields in
$2+1$ Minkowski space-time, interacting with a magnetic background
field perpendicular to the spatial plane, and evaluate the vacuum
expectation values of the energy and fermion density.

Section
\ref{Sect3} contains the generalities of the same theory in Euclidean $3$-dimensional space, and presents the eigenvalues of
the corresponding Dirac operator. From such eigenvalues, the
partition function is evaluated in section \ref{Sect4}.

Section
\ref{Sect5} contains the resulting free energy and mean particle
density at finite temperature, and a study of their zero-temperature limit. In the same section, we
perform an adequate Lorentz boost in order to consider the problem
of fermions interacting with crossed electric and magnetic fields,
and obtain the Hall conductivity.

Some final comments about the case in which the chemical potential $\mu$ coincides with an energy level are presented in section \ref{Sect6}. In the same section, we discuss the role that experimental results on the quantum Hall effect can play, as an arena for testing the physical relevance of the phase of the determinant or, equivalently, of the multiplicative anomaly, in view of our discrepancies with reference \cite{gusynhall}.

A very sketchy presentation of the present study can be found in \cite{proceednos}.

\section{Zero-temperature problem}\label{Sect2}

We will use the metric $(-,+,+)\,,$
natural units $\hbar=c=1$, and choose the following representation
for the Dirac matrices: namely,
\beq
\gc=i\sigma_3,\qquad\gu=\sigma_2\qquad {\rm and}\qquad
\gd=\sigma_1.\label{gammam}
\eeq
The Hamiltonian can be determined from the
solutions of the Dirac equation $(\dirac)\Psi=0$, where $-e$ is the negative charge of the electron. In the Landau gauge $A=(0,0,Bx)\,,$
with $B>0$. Thus, after setting
$\Psi(t,x,y)=e^{-iEt}\psi(x,y)$, we get the Hamiltonian $H=i\sigma_1
\partial_x-i\sigma_2
\partial_y+\sigma_2 eB x$.
The corresponding eigenvalues can be determined from
\beq
\left(\begin{array}{cc}
  -E & i\partial_x-\partial_y-i e B x \\
  i\partial_x+\partial_y+i e B x  & -E \\
\end{array}\right) \psi(x,y)\ =\ 0\,.
\eeq

In order to solve the above equation, we take
\beq
\psi_k(x,y)=\left(\begin{array}{c}
  \varphi_k(x,y) \\
  \chi_k(x,y) \\
\end{array}\right) =\frac{1}{\sqrt{2\pi}}
\left(\begin{array}{c}
e^{iky} \varphi_{k} (x)\\
e^{iky} \chi_{k} (x) \\
\end{array}\right)\,.\eeq

This leads to the following system of first order equations \beq
\nn (i\partial_x -i k -i e B
x)\chi_{k}&=&E\varphi_{k}\\(i\partial_x +i k + i e B
x)\varphi_{k}&=& E\chi_{k} \,. \eeq

We first solve for the zero mode $E_0=0$. In this case, both
equations decouple and, after imposing that the eigenfunctions be
well-behaved for $x \rightarrow \pm \infty$ we obtain, up to
normalization
\beq \psi_k (x)=\left(\begin{array}{c}
  (e B/\pi)^{1/4} \exp\left\{-\frac{1}{2}\,e B (x+k/e B)^2\right\}\\
0\\
\end{array}\right) \,.
\eeq

In the case $E\neq 0$, the normalized eigenfunctions which are
well-behaved for all values of $x$ can be written in terms of the
Hermite's polynomials $H_n(u)$ \cite{gradshteyn} as follows

\beq
\psi_{k,n}^\pm (x) &=&
\frac{(e B/\pi)^{1/4}}{\sqrt{n!\,2^{n+1}}}\,
\exp\left\{-\frac{1}{2}\,e B(x+{k}/{e B})^2\right\}\\\nonumber
&\times&
\left(\begin{array}{l}
\pm \,H_{n}\left(\sqrt{e B}\,(x+{k}/{e B})\right)\\
i\sqrt{2n}\ H_{n-1}\left(\sqrt{e B}\,(x+{k}/{e B})\right)\\
\end{array}\right) \,,
\eeq
with corresponding eigenvalues
\beq
E_n= \pm \sqrt{2n e B}\,,\qquad\quad
n=1,..., \infty\,. \label{espectro}
\eeq
We notice that each eigenvalue exhibits the well known Landau's
degeneracy per unit area: namely,
\beq
\Delta_L\ =\ \frac{e B}{2\pi}\,.
\label{deg}\eeq

Had we chosen the other non-equivalent representation of
the gamma matrices in $2+1$ dimensions, exactly the same spectrum would have been obtained obtained, the only difference being the chirality of eigenfunctions.

The vacuum expectation value of the energy per unit area, defined
through a zeta function regularization (see, for example,
\cite{elizalde}, and references therein), is given by \beq  E_C =
-\left.\frac{\Delta_L}{2} \sum_{E_n\neq 0}
|E_n|^{-s}\right\rfloor_{s=-1}\,.\label{ecas}\eeq

In the present case, we have ($\alpha$ is an arbitrary parameter
with mass dimension, introduced to render the complex powers
dimensionless)

\beq \fl E_C(B) = -\left.\frac{\Delta_L\alpha}{2} 2 \sum_{n=1}^{\infty}
\left(\frac{\sqrt{2n e B}}{\alpha}\right)^{-s}\right\rfloor_{s=-1}=
-\Delta_L \sqrt{2 e B}\, \zeta_R \left(-\frac{1}{2}\right)\,.\eeq

Always in the zeta-function regularization framework, the fermion
number is \cite{ns} \beq \nn N(B) &=& -\left.\frac{\Delta_L}{2}
\left(\sum_{E_n > 0} |E_n|^{-s}-\sum_{E_n < 0}
|E_n|^{-s}\right)\right\rfloor_{s=0}+N_0\,,\label{numero}\eeq
where $N_0$ is the contribution coming from zero modes.

In our case, the nonvanishing spectrum is symmetric. So, only the
zero mode, which is charge self-conjugate, contributes. This gives
as a result \cite{ns}
 \beq N(B) =\pm \frac{\Delta_L}{2}\,.\label{nf}\eeq
Or, equivalently, for the vacuum expectation value of the charge density
\beq j^0(B) =\mp e\frac{\Delta_L}{2}\,.\eeq

The sign indetermination is a natural consequence of the twofold vacuum degeneracy.

\section{The theory at finite temperature with chemical potential}\label{Sect3}

In order to study the effect of temperature, we go to Euclidean
space, with the metric $(+,+,+)\,.$ To this end, we take the Euclidean
gamma matrices to be $\gce=i\gc=-\sigma_3$, $\gue=\gu=\sigma_2$,
$\gamma_2=\gamma_M^2=\sigma_1$. We will follow \cite{actor} in
introducing the chemical potential as an imaginary $A_0=-i\frac{\mu}{e}$ in
Euclidean space. Thus,  the partition function in the
grand-canonical ensemble is given by
\beq
\ln{Z}=\ln\,
{\rm det}(\dirac)\,.
\label{??}
\eeq

In order to evaluate the partition function in the zeta
 regularization  approach \cite{dowker}, we first determine the eigenfunctions,
 and the corresponding eigenvalues, of the Dirac operator, in the
 same gauge used in the previous section, i.e, we solve
\beq [-i\sigma_3(\partial_{\tau}+\mu)+
 i\sigma_2
\partial_x+\sigma_1
(i\partial_y-e B x)]\Psi=\omega \Psi\,,\eeq or, after writing
$\Psi({\tau},x,y)=\left(\begin{array}{c}
    \Phi({\tau},x,y) \\
    \Xi({\tau},x,y) \\
  \end{array}\right)$,
  \beq \fl \left(\!\!\matrix{
  -i(\partial_{\tau}+\mu)&\partial_x+i\partial_y-e Bx \cr
  -\partial_x+i\partial_y-eBx& i(\partial_{\tau}+\mu)} \right)\!\!\left(\begin{array}{c}
    \Phi({\tau},x,y) \\
    \Xi({\tau},x,y) \\
  \end{array}\right)=\omega\left(\begin{array}{c}
    \Phi({\tau},x,y) \\
    \Xi({\tau},x,y) \\
  \end{array}\right)\,.\eeq

  In order to satisfy antiperiodic boundary conditions in the
  ${\tau}$ direction, we propose
  \beq
  \Psi_{k,l}({\tau},x,y)=\frac{e^{i\lambda_l {\tau}}e^{iky}}{\sqrt{2\pi\beta}}\, \psi_{k,l} (x)\,,\eeq
  with
  \beq
  \lambda_l=(2l+1)\frac{\pi}{\beta}\,,\label{lambda}\eeq
where $\beta=\frac{1}{T}$ is the inverse temperature.

  After doing so, and writing \[
\psi_{k,l} (x)=\left(\begin{array}{c}
  \varphi_{k,l} (x) \\
  \chi_{k,l} (x) \\
\end{array}\right) \,,
\] we have, for each
  $k,l$,
  \beq \nn (\partial_x-k-e Bx)\chi_{k,l}&=&(\omega-\tilde \lambda_l)
\varphi_{k,l}
\\(-\partial_x-k-e Bx)\varphi_{k,l}&=&(\omega+\tilde \lambda_l) \chi_{k,l}
\,,\label{diffeq}\eeq where $\tilde \lambda_l=\lambda_l-i\mu$.

There are two types of eigenvalues and corresponding
eigenfunctions

\bigskip

1) $\omega_l=\tilde \lambda_l$, with $l=-\infty,...,\infty$, and
corresponding normalized eigenfunctions \beq \psi_{k,l}
(x)=\left(\begin{array}{c}
  \left(\frac{eB}{\pi}\right)^{\frac14} e^{-\frac{e B}{2} (x+\frac{k}{e B})^2} \\0
  \\
\end{array}\right) \,.\eeq

Note that these eigenvalues are not square roots of the eigenvalues of the squared operator. They will eventually lead to a ``spectral asymmetry" \footnote{Here, quotation marks are due to the fact that, the Dirac operator not being self-adjoint, we have a complex spectrum.} and, thus, to a phase of the determinant, which will be studied in detail in the next section.

\bigskip

2) $\omega_{l,n}=\pm \sqrt{{\tilde \lambda_l}^2+2n e B}$, with
$n=1,...,\infty$, $l=-\infty,...,\infty$, and corresponding
normalized eigenfunctions \beq \fl \psi_{k,l,n} (x)= A_{k,l,n}
\left(\begin{array}{l}
-\frac{(\omega_{l,n}+\tilde \lambda_{l})}{2n\sqrt{e B}}
e^{-\frac{e B}{2}(x+\frac{k}{e B})^2}
 H_{n}(\sqrt{e B}(x+\frac{k}{e B}))\\ \\ e^{-\frac{e B}{2}(x+\frac{k}{e B})^2} H_{n-1} (\sqrt{e B}(x+\frac{k}{e B}))
\\
\end{array}\right) \,,\eeq
where
\[ A_{k,l,n}=\left(\frac{e B}{\pi}\right)^{\frac14}
\frac{2^{\frac{1-n}{2}}}{[(n-1)!\,]^{\frac12}}
\left[\frac{2ne B}{2ne B+|\omega_{l,n}+\tilde
\lambda_{l}|^2}\right]^{\frac12} \,.\]

\bigskip

In all cases, the degeneracy per unit area is again given by $\Delta_L$ in equation (\ref{deg}).
Choosing the other nonequivalent representation of the gamma matrices leads to a change in the eigenvalues of type 1), which are replaced by $\omega_l=-\tilde \lambda_l$. However, as will be discussed below, this doesn't lead to a change in our physical predictions.

\section{Evaluation of the partition function at finite temperature and chemical potential}\label{Sect4}

The partition function, in the zeta regularization scheme \cite{dowker}, is
given by \beq \left.\log{{\cal
Z}}=-\frac{d}{ds}\right\rfloor_{s=0}\,\zeta(s,\frac{\dirac}{\alpha})\,.
\label{partfunc}\eeq As in the previous section, $\alpha$ is a
parameter with mass dimension, introduced to render the
$\zeta$-function dimensionless.

We must consider two contributions to $\log{{\cal Z}}$, respectively coming from eigenvalues of type 1) and 2) in the previous section, i.e.,

\beq \Delta_1(\mu)=\left.
-\frac{d}{ds}\right\rfloor_{s=0}{\zeta}_1
(s,\mu)\,,\label{del2mu}\eeq where \beq  {\zeta}_1 (s,\mu)=\Delta_L
\sum_{l=-\infty }^{\infty}\left[ (2l+1)\frac{\pi}{\alpha
\beta}-i\frac{\mu}{\alpha}\right]^{-s}\,,\label{z2mu}\eeq and \beq
\Delta_2(\mu, B)=\left. -\frac{d}{ds}\right\rfloor_{s=0}{\zeta}_2
(s,\mu, B)\,,\label{del1mu}\eeq where \beq \fl {\zeta}_2
(s,\mu,B)=(1+(-1)^{-s})\, \Delta_L \!\!\!\!\!\!\!\sum_{\begin{array}{c}
  n=1 \\
  l=-\infty \\
\end{array}}^{\infty}\!\!\!\!\!\left[\frac{2ne B}{\alpha^2}+
 {\left((2l+1)\frac{\pi}{\alpha
\beta}-i\frac{\mu}{\alpha}\right)}^2\right]^{-\frac{s}{2}}\!\!.\label{z1mu}\eeq

The contribution $\Delta_1(\mu)$ can be evaluated at once for the
whole $\mu$-range. The analytic extension of ${\zeta}_1 (s,\mu)$
can be achieved as follows (for a similar calculation, see \cite{bagnos})

\beq \fl \nn {\zeta}_1 (s,\mu)&=&\Delta_L \sum_{l=-\infty
}^{\infty}\left[ (2l+1)\frac{\pi}{\alpha
\beta}-i\frac{\mu}{\alpha}\right]^{-s}\\
\fl \nn &=& \Delta_L
\left(\frac{2\pi}{\alpha\beta}\right)^{-s}\left[\sum_{l=0
}^{\infty}\left[
(l+\frac12)-i\frac{\mu\beta}{2\pi}\right]^{-s}+\sum_{l=0
}^{\infty}\left[
-(l+\frac12)-i\frac{\mu\beta}{2\pi}\right]^{-s}\right]\\\fl &=&
 \Delta_L \left(\frac{2\pi}{\alpha\beta
}\right)^{-s}\left[ \zeta_H \left(s,\frac12-\frac{i\mu
\beta}{2\pi}\right)+\sum_{l=0 }^{\infty}\left[
-(l+\frac12)-i\frac{\mu\beta}{2\pi}\right]^{-s}\right]\,.\eeq

Now, in order to write the second term as a Hurwitz zeta, we must
relate the eigenvalues with negative real part to those with
positive one without, in so doing, going through zeros in the
argument of the power. Otherwise stated, we must select a cut in
the complex $\omega$ plane \cite{ecz}. This requirement determines
a definite value of $(-1)^{-s}$, i.e., $(-1)^{-s}=e^{i\pi sign
(\mu)s}$. Taking this into account, we finally have \beq\fl
\zeta_1(s,\mu)=\Delta_L \left(\frac{2\pi}{\beta
\alpha}\right)^{-s}\left[\zeta_H \!\left(\!s,\frac12-\frac{i\mu
\beta}{2\pi}\right)+e^{i\pi sign(\mu)s}\zeta_H
\!\left(\!s,\frac12+\frac{i\mu \beta}{2\pi}\right)\right]
\label{ext1mu}.\eeq

From this last expression, the contribution $\Delta_1(\mu)$ to
$\log{{\cal Z}}$ can be obtained. It is given by
\beq\fl\nn
\Delta_1(\mu) &=& -\Delta_L\left[\zeta_H^{\prime}\!\left(\!0, \frac12-
\frac{i\mu \beta}{2\pi}\right)+\zeta_H^{\prime}\!\left(\!0, \frac12
+\frac{i\mu \beta}{2\pi}\right)+i\pi sign(\mu)\zeta_H \!\left(\!0,
\frac12 +\frac{i\mu \beta}{2\pi}\right)\right]\\
\fl &=&\Delta_L \left\{\log{\left(2\cosh{\left(\frac{\mu
\beta}{2}\right)}\right)}-\frac{|\mu|\beta}{2}\right\}\,.\label{delta1}\eeq

As commented in advance, the other nonequivalent representation of the gamma matrices leads to the same result for this contribution. In fact, even though this part of the spectrum changes sign, such change is compensated by the selection of the cut in the $\omega$-plane. In this case, one has
 \beq\fl \nn
\zeta_1(s,\mu)=\Delta_L \left(\frac{2\pi}{\beta
\alpha}\right)^{-s}\left[\zeta_H \!\left(\!s,\frac12+\frac{i\mu
\beta}{2\pi}\right)+e^{-i\pi sign(\mu)s}\zeta_H
\!\left(\!s,\frac12-\frac{i\mu \beta}{2\pi}\right)\right]
,\eeq
which also leads to (\ref{delta1}).

\bigskip

The analytic extension of ${\zeta}_2(s,\mu,B)$ requires a separate
consideration of different $\mu$ ranges. We study in detail three of
these cases. The generalization to arbitrary $\mu$-ranges will be evident from these results.

\subsection{$ {\mu}^2 < 2e B$}

\beq \fl {\zeta}_2(s,\mu,B)=(1+(-1)^{-s})\, \Delta_L
\!\!\!\!\!\!\!\sum_{\begin{array}{c}
  n=1 \\
 l=-\infty \\
\end{array}}^{\infty}\!\!\!\!\left[\frac{2ne B}{\alpha^2}+
{\left((2l+1)\frac{\pi}{\alpha
\beta}-i\frac{\mu}{\alpha}\right)}^2\right]^{-\frac{s}{2}}\,.\label{z2mumenor}\eeq

Making use of the Mellin transform, this can be written as
\beq\fl
{\zeta}_2(s,\mu,B)=\frac{(1+(-1)^{-s})\,\Delta_L}{\Gamma(\frac{s}{2})}\int_0^{\infty}dt\,
t^{\frac{s}{2}-1}\!\!\!\!\sum_{\begin{array}{c}
  n=1 \\
  l=-\infty \\\end{array}}^{\infty}\!\!\!\!e^{-t\left[\frac{2ne B}{\alpha^2}+
{\left((2l+1)\frac{\pi}{\alpha
\beta}-i\frac{\mu}{\alpha}\right)}^2\right]}\eeq
or, equivalently,

 \beq \nn {\zeta}_2
(s,\mu,B)&=&\frac{(1+(-1)^{-s})\,
\Delta_L}{\Gamma(\frac{s}{2})}\sum_{n=1}^{\infty}\int_0^{\infty}dt\,
t^{\frac{s}{2}-1}e^{-t
\left[\frac{2ne B}{\alpha^2}+\left(\frac{\pi}{\alpha
\beta}-\frac{i\mu}{\alpha
}\right)^2\right]}\\&\times&\Theta_3\left(\frac{-2t}{\alpha
\beta}\left(\frac{\pi}{\alpha
\beta}-\frac{i\mu}{\alpha}\right),\frac{4\pi t}{(\alpha
\beta)^2}\right)\,,\eeq where we have used the definition of the
Jacobi theta function \beq \Theta_3
(z,x)=\sum_{l=-\infty}^{\infty}e^{-\pi x l^2} e^{2\pi z
l}\,.\label{th3}\eeq

To proceed, we will use the inversion formula for the Jacobi
function \beq \Theta_3(z,x)=\frac{1}{\sqrt{x}}e^{(\frac{\pi
z^2}{x})}\Theta_3\left(\frac{z}{ix},\frac{1}{x}\right)\,,\eeq
thus getting
 \beq \nn {\zeta}_2(s,\mu,B)&=&\frac{(1+(-1)^{-s})\, \Delta_L \alpha \beta}{2\sqrt{\pi}\Gamma(\frac{s}{2})}\sum_{n=1}^{\infty}\int_0^{\infty}dt\,
t^{\frac{s-1}{2}-1}e^{-t \frac{2ne B}{\alpha^2}}\\&\times&\Theta_3\left(\frac{i}{2}+\frac{\mu \beta}{2\pi},\frac{(\alpha \beta)^2}{4\pi t}\right)\,.\eeq

Applying once more the definition (\ref{th3}), we have
 \beq\fl \nn {\zeta}_2(s,\mu,B)&=&\frac{(1+(-1)^{-s})\, \Delta_L \alpha \beta}{2\sqrt{\pi}\Gamma(\frac{s}{2})}\left\{\int_0^{\infty}dt\,
t^{\frac{s-1}{2}-1}\sum_{n=1}^{\infty}e^{-t \frac{2ne B}{\alpha^2}}\right.\\ \fl&+&\left.  2\int_0^{\infty}\!dt\,
t^{\frac{s-1}{2}-1}\sum_{n,l=1}^{\infty}(-1)^l \cosh{(\mu \beta
l)}e^{-t \frac{2ne B}{\alpha^2}-\frac{{(\alpha \beta l)}^2}{4t}}\right\}\,.\eeq

After performing the integrals \cite{gradshteyn}, we obtain \beq\fl
\nn {\zeta}_2(s,\mu,B)&=&\frac{(1+(-1)^{-s}) \Delta_L \alpha
\beta}{2\sqrt{\pi}\Gamma(\frac{s}{2})}
\left[\Gamma\left(\frac{s-1}{2}\right)\left(\frac{2e B}{
\alpha^2}\right)^{\frac{1-s}{2}}\zeta_R
\left(\frac{s-1}{2}\right)\right.\\\fl &+&\left.4\sum_{n,l=1}^{\infty}(-1)^l
\left(\frac{l^2 \alpha^4
\beta^2}{8ne B}\right)^{\frac{s-1}{4}}\!\!\cosh{(\mu \beta
l)}K_{\frac{s-1}{2}}\left(\!\sqrt{2ne B\beta^2 l^2}\right)\!\right] \eeq
or, making the simple pole of the $\Gamma$ function explicit \beq
\nn\fl {\zeta}_2(s,\mu,B)&=&\frac{(1+(-1)^{-s}) \Delta_L \alpha \beta
s}{4\sqrt{\pi}\Gamma(\frac{s+2}{2})}
\left[\Gamma\left(\frac{s-1}{2}\right)\left(\frac{2e B}{
\alpha^2}\right)^{\frac{1-s}{2}}\zeta_R
\left(\frac{s-1}{2}\right)\right.\\\fl &+&\left.4\sum_{n,l=1}^{\infty}\!\!(-1)^l
\left(\frac{l^2 \alpha^4
\beta^2}{8ne B}\right)^{\frac{s-1}{4}}\!\!\!\cosh{(\mu \beta
l)}K_{\frac{s-1}{2}}\left(\!\sqrt{2ne B\beta^2 l^2}\right)\right]\,.
\label{ext2mu}\eeq

From this expression, the contribution $\Delta_2$ to the partition
function can be readily obtained, since the factor accompanying $s$
is finite at $s=0$ \beq\fl \nn {\Delta}_2(\mu,B) &=&\frac{- \Delta_L
\beta }{2\sqrt{\pi}}
\left[\Gamma\left(-\frac{1}{2}\right)\sqrt{2e B}\zeta_R
\left(-\frac{1}{2}\right)
\right.\\\fl&+&\left.4\sum_{n,l=1}^{\infty}(-1)^l \left(\frac{l^2
\beta^2}{8ne B}\right)^{-\frac{1}{4}}\!\!\cosh{(\mu \beta
l)}K_{-\frac{1}{2}}\left(\!\sqrt{2ne B\beta^2 l^2}\right)\right]\,.
\eeq

After using that
\[K_{-\frac{1}{2}}(x)=\sqrt{\frac{\pi}{2x}}e^{-x}\,,\]
it can be written as \beq \nn {\Delta}_2(\mu,B) &=& \Delta_L \beta
\left[\sqrt{2e B}\zeta_R
\left(-\frac{1}{2}\right)\right.\\&-&\left.\frac{2}{\beta}\sum_{n,l=1}^{\infty}
\frac{(-1)^l}{l} \cosh{(\mu \beta l)}e^{-\sqrt{2ne B}\beta
l}\right]\,. \eeq

The sum over $l$ can be explicitly performed, to obtain \beq\fl \nn
{\Delta}_2(\mu,B) &=& \Delta_L \beta \left[\sqrt{2e B}\zeta_R
\left(-\frac{1}{2}\right)\right.\\\fl&+&\left.\frac{1}{\beta}
\sum_{n=1}^{\infty}
\log{\left(1+ e^{-2\sqrt{2ne B}\beta}+2\cosh{(\mu \beta
)}e^{-\sqrt{2ne B}\beta}\right)}\right]\,. \label{delta2}\eeq

Finally, adding the contributions given by equations
(\ref{delta1}) and (\ref{delta2}) we obtain, for the partition
function in the range ${\mu}^2 \leq 2e B$ \beq \nn\log{Z}&=&\Delta_L
\left\{\log{\left(2\cosh{\left(\frac{\mu
\beta}{2}\right)}\right)}-\frac{|\mu|\beta}{2}\right.+
\beta \sqrt{2e B}\zeta_R \left(-\frac{1}{2}\right)\\&+&
\left.\sum_{n=1}^{\infty} \log{\left(1+
e^{-2\sqrt{2ne B}\beta}+2\cosh{(\mu \beta
)}e^{-\sqrt{2ne B}\beta}\right)}\right\}\,.\label{zmenor}\eeq

\subsection{$2e B<{\mu}^2 < 4e B$}

As before, we have \beq\fl {\zeta}_2 (s,\mu,B)=(1+(-1)^{-s})\, \Delta_L
\sum_{\begin{array}{c}
  n=1 \\
  l=-\infty \\
\end{array}}^{\infty}\left[\frac{2ne B}{\alpha^2}+
{\left((2l+1)\frac{\pi}{\alpha
\beta}-i\frac{\mu}{\alpha}\right)}^2\right]^{-\frac{s}{2}}\!\!\,.\eeq

However, in this range of $\mu$, the contribution due to $n=1$ is given by
\beq {\Delta}_2^{n=1}(\mu,B)=\left.
-\frac{d}{ds}\right\rfloor_{s=0}{\zeta}_2^{n=1} (s,\mu,B),\label{ddel2mu}\eeq
where

\beq \fl {\zeta}_2^{n=1} (s,\mu,B)=(1+(-1)^{-s})\, \Delta_L \sum_{
  l=-\infty }^{\infty}\left[\frac{2e B}{\alpha^2}+
{\left((2l+1)\frac{\pi}{\alpha
\beta}-i\frac{\mu}{\alpha}\right)}^2\right]^{-\frac{s}{2}}\,.\eeq

The analytic extension of this expression must be performed in a different way. In fact, the expression cannot be written in terms of a unique Mellin transform, since its real part is not always positive (note, in connection with this that, for $n=1$, equation
(\ref{ext2mu}) diverges). Instead, it can be evaluated as follows
\beq\fl \nn {\zeta}_2^{n=1} (s,\mu,B)&=&\frac{(1+(-1)^{-s})}{\alpha^{-s}}\, \Delta_L \sum_{
  l=0 }^{\infty}\left[2e B+
{\left((2l+1)\frac{\pi}{
\beta}-i\mu\right)}^2\right]^{-\frac{s}{2}} \\\fl &+&\nn \mu \rightarrow -\mu\\ \fl\nn &=&\frac{(1+(-1)^{-s})}{\alpha^{-s}}\, \Delta_L \sum_{
  l=0 }^{\infty}\left\{\left[i\sqrt{2e B}+
(2l+1)\frac{\pi}{
\beta}-i\mu\right]^{-\frac{s}{2}}\right.\\\fl &\times& \left.\left[-i\sqrt{2e B}+
(2l+1)\frac{\pi}{
\beta}-i\mu\right]^{-\frac{s}{2}}\right\}+\mu \rightarrow -\mu\,.\eeq

This can be written as a product of two Mellin transforms

\beq \fl \nn {\zeta}_2^{n=1} (s,\mu,B)&=&\frac{(1+(-1)^{-s})}{\alpha^{-s}[\Gamma(\frac{s}{2})]^2}\, \Delta_L \sum_{l=0 }^{\infty}\int_0^{\infty}dt\,
t^{\frac{s}{2}-1}e^{-\left[
(2l+1)\frac{\pi}{
\beta}-i\mu+i\sqrt{2e B}\right]t}\\\fl&\times&\int_0^{\infty}dz\,
z^{\frac{s}{2}-1}e^{-\left[
(2l+1)\frac{\pi}{
\beta}-i\mu-i\sqrt{2e B}\right]z}+\mu \rightarrow -\mu\eeq
or, after a change of variables
\beq \nn {\zeta}_2^{n=1} (s,\mu,B)&=&\frac{(1+(-1)^{-s})}{2\alpha^{-s}[\Gamma(\frac{s}{2})]^2}\, \Delta_L \sum_{l=0 }^{\infty}\int_0^{\infty}dz\,\int_{-z}^{z}dt\,
\left[\frac{z^2 -t^2}{4}\right]^{\frac{s}{2}-1} \\&\times&e^{-\left[
(2l+1)\frac{\pi}{
\beta}-i\mu\right]z}e^{
-i\sqrt{2e B}t}+\mu \rightarrow -\mu \,.\eeq

Now, calling $x=\frac{t}{z}$, one has
\beq\fl {\zeta}_2^{n=1} (s,\mu,B)&=&\nn \frac{(1+(-1)^{-s})}{2\alpha^{-s}[\Gamma(\frac{s}{2})]^2}\, \Delta_L \sum_{l=0 }^{\infty}\int_0^{\infty}dz\,z\left[\frac{z^2}{4}\right]^{\frac{s}{2}-1}
e^{-\left[
(2l+1)\frac{\pi}{
\beta}-i\mu\right]z}\\\fl &\times&\int_{-1}^{1}dx\,
\left[1-x^2\right]^{\frac{s}{2}-1}e^{
-i\sqrt{2e B}zx}+\mu \rightarrow -\mu \,.\eeq

Finally, the $x$-integral and the sum of the geometric series can be performed to obtain

\beq \nn {\zeta}_2^{n=1} (s,\mu,B)&=&\frac{(1+(-1)^{-s})\sqrt{\pi}}{2\alpha^{-s}\Gamma(\frac{s}{2})}\, \Delta_L \left(2\sqrt{2e B}\right)^{\frac{1-s}{2}}\\&\times&\int_0^{\infty}dz\,
z^{\frac{s-1}{2}}
J_{\frac{s-1}{2}}(\sqrt{2e B}z)\frac{e^{i\mu z}}{\sinh{(\frac{\pi z}{\beta})}}+\mu \rightarrow -\mu \,.\eeq

Now, the integral in this expression diverges at $z=0$. In order to isolate such divergence, we add and subtract the first term in the series expansion of the Bessel function, thus getting the following two pieces

\beq\fl \nn {\zeta}_{2,(1)}^{n=1} (s,\mu,B)&=&\frac{(1+(-1)^{-s})\sqrt{\pi}s}{4\alpha^{-s}\Gamma(\frac{s}{2}+1)}\, \Delta_L \left(2\sqrt{2e B}\right)^{\frac{1-s}{2}}\\\fl\nn&\times&\int_0^{\infty}dz\,
z^{\frac{s-1}{2}}
\left[J_{\frac{s-1}{2}}(\sqrt{2e B}z)-
\frac{\left(\frac{\sqrt{2e B}z}{2}\right)^{\frac{s-1}{2}}}
{\Gamma\left(\frac{s+1}{2}\right)}\right]\frac{e^{i\mu z}}{\sinh{(\frac{\pi z}{\beta})}}\\\fl &+&\mu \rightarrow -\mu \,,\label{zeta21}\eeq
and
\beq \fl {\zeta}_{2,(2)}^{n=1} (s,\mu,B)=\frac{(1+(-1)^{-s})\sqrt{\pi}}{2^s \alpha^{-s}\Gamma(\frac{s}{2})\Gamma(\frac{s+1}{2})}\, \Delta_L \int_0^{\infty}dz\,
z^{{s-1}}\frac{e^{i\mu z}}{\sinh{(\frac{\pi z}{\beta})}}+\mu \rightarrow -\mu \,.\label{zeta22}\eeq

The contribution of equation (\ref{zeta21}) to the partition function, defined as in equation (\ref{ddel2mu}) can be easily evaluated by noticing that the factor multiplying $s$ is finite at $s=0$. Thus, one has
\beq\fl {\Delta}_{2,(1)}^{n=1}(\mu,B)=-\Delta_L \int_0^{\infty}\!dz\,z^{-1} \left[\cos{(\sqrt{2e B}z)}-1\right]\frac{e^{i\mu z}}{\sinh{(\frac{\pi z}{\beta})}}+\mu \rightarrow -\mu \,,\label{inter}\eeq
where we have used that $J_{-\frac{1}{2}}(\sqrt{2e B}z)=\sqrt{\frac{2}{\pi \sqrt{2e B}z}}\cos{(\sqrt{2e B}z)}$. Now, in the term with $\mu\rightarrow -\mu$, one can change $z\rightarrow -z$ to obtain
\beq {\Delta}_{2,(1)}^{n=1}(\mu,B)=-\Delta_L\int_{-\infty}^{\infty}dz\,z^{-1} \left[\cos{(\sqrt{2e B}z)}-1\right]\frac{e^{i\mu z}}{\sinh{(\frac{\pi z}{\beta})}} \,.\eeq
This last integral is easy to evaluate in the complex plane, by carefully taking into account the sign of $\mu$, as well as well as the fact that $2e B<{\mu}^2$ in closing the integration path, to obtain
\beq\fl {\Delta}_{2,(1)}^{n=1}(\mu,B)=-2\Delta_L\sum_{l=1}^{\infty}
\left[\frac{(-1)^l}{l}\cosh{(\sqrt{2e B} \beta l
)}e^{-|\mu|\beta l}+\frac{(-1)^{l+1}}{l}e^{-|\mu|\beta l}\right]\eeq
or, after summing the series
 \beq \nn {\Delta}_{2,(1)}^{n=1}(\mu,B)&=&\Delta_L\left\{\log{\left(1+e^{-2|\mu|\beta}
+2\cosh{(\sqrt{2e B}\beta)}e^{-|\mu|\beta}\right)}\right.\\
&+&\left.
|\mu|\beta -
2\log\left(2\cosh{(\frac{\mu\beta}{2}})\right)\right\}\,.\label{del21may}\eeq

In order to obtain the contribution coming from (\ref{zeta22}), the integral can be evaluated for $\Re s>1$, which gives
\beq \nn {\zeta}_{2,(2)}^{n=1} (s,\mu,B)&=&\frac{(1+(-1)^{-s})\Gamma(s) \sqrt{\pi}{(\alpha \beta)}^{s} \Delta_L}{{(2\pi)}^{s}2^{s-1}\Gamma(\frac{s}{2})\Gamma(\frac{s+1}{2})} \\&\times&\left[\zeta_H (s,\frac12 (1-\frac{i\mu\beta}{\pi}))+\zeta_H (s,\frac12 (1+\frac{i\mu\beta}{\pi}))\right] \,,\eeq
where $\zeta_H(s,x)$ is the Hurwitz zeta function. The contribution to the partition function can now be evaluated by using that $\zeta_H (0,\frac12 (1-\frac{i\mu\beta}{\pi}))+\zeta_H (0,\frac12 (1+\frac{i\mu\beta}{\pi})=0$ and the well known value of $-\frac{d}{ds}\rfloor_{s=0}\zeta_H(s,x)$ \cite{gradshteyn}, to obtain
\beq {\Delta}_{2,(2)}^{n=1}(\mu,B)=2\Delta_L\log(2\cosh{(\frac{\mu\beta}{2}}))
\,.\label{del22may}\eeq

Summing up the contributions in equations (\ref{delta1}), (\ref{del21may}) and (\ref{del22may}), as well as the contribution coming from $n\geq 2$, evaluated as in the previous subsection, one gets for the partition function

\beq\fl \nn \log{Z}&=&\Delta_L \left\{\log{\left(2\cosh{\left(\frac{\mu
\beta}{2}\right)}\right)}+\frac{|\mu|\beta}{2}\right.\\\fl&+& \nn \log{\left(1+
e^{-2|\mu|\beta}+2\cosh{(\sqrt{2e B}\beta)}e^{-|\mu|\beta}\right)}
+\beta \sqrt{2e B}\left(\zeta_R \!\left(\!-\frac{1}{2}\right)-1\right)\\\fl &+&\left.\sum_{n=2}^{\infty} \log{\left(1+
e^{-2\sqrt{2ne B}\beta}+2\cosh{(\mu \beta
)}e^{-\sqrt{2ne B}\beta}\right)}\right\}
\label{zmayor}\,.\eeq

At first sight, this result looks different from the one corresponding to $\mu^2<2e B$ (equation (\ref{zmenor})). However, it is easy to see that both expressions coincide. The only difference is that (\ref{zmayor}) explicitly isolates the zero-temperature behavior from finite-temperature corrections for this range of $\mu$.

\subsection{${\mu}^2 =2e B$}

In this case, the analytical extension can be performed, exactly as in the previous subsection, up to equation (\ref{inter}). Then, the term with $\mu\rightarrow -\mu$ can be explicitly summed, which gives
\beq\fl {\Delta}_{2,(1)}^{n=1}(|\mu|=\sqrt{2e B},B)=-\Delta_L \int_0^{\infty}\!dz\,z^{-1} \left[\cos{(\sqrt{2e B}z)}-1\right]\frac{2\cos{(\sqrt{2e B}z)}}{\sinh{(\frac{\pi z}{\beta})}}\,.\eeq

This integral can be found in \cite{gradshteyn}, and it gives as a result
\beq \fl {\Delta}_{2,(1)}^{n=1}(|\mu|=\sqrt{2e B},B)=\Delta_L\left[\log{\left(\cosh{\left(\sqrt{2e B}\beta\right)}\right)}\!\!-2\log{\left(\cosh{\left(\frac{\sqrt{2e B}\beta}{2}\right)}\right)}\right]\,.\eeq

The contribution ${\Delta}_{2,(2)}^{n=1}(|\mu|=\sqrt{2e B},B)$ can be evaluated exactly as in the previous subsection. It is given by

\beq {\Delta}_{2,(2)}^{n=1}(|\mu|=\sqrt{2e B},B)=2\Delta_L\log(2\cosh{(\frac{\sqrt{2e B}\beta}{2}}))
\,.\eeq

These two contributions, together with those coming from $n\geq 2$ and from equation (\ref{delta1}), finally give for the partition function at this particular value of $\mu$ the same result as (\ref{zmenor}) or (\ref{zmayor}) evaluated at ${\mu}^2 =2e B$.

\bigskip

Thus, for any range of $\mu$, one has
\beq \nn\log{Z}&=&\Delta_L
\left\{\log{\left(2\cosh{\left(\frac{\mu
\beta}{2}\right)}\right)}-\frac{|\mu|\beta}{2}\right.+
\beta \sqrt{2e B}\zeta_R \left(-\frac{1}{2}\right)\\&+&
\left.\sum_{n=1}^{\infty} \log{\left(1+
e^{-2\sqrt{2ne B}\beta}+2\cosh{(\mu \beta
)}e^{-\sqrt{2ne B}\beta}\right)}\right\}\,,\label{partfunct}\eeq
which is continuous, even when $\mu$ coincides with an energy level.

When this calculation was finished, it was pointed to us that, in reference \cite{gusynNJL}, the partition function for Dirac fields interacting not only with the background magnetic field, but also among themselves through a Nambu-Jona-Lasinio term,  had been obtained through a different regularization (see also \cite{klimenko}; for a calculation of parity-breaking corrections, see \cite{rossini}). However, in the absence of this last interaction, the result in such reference differs from the present one for two reasons: In the first place, the authors of \cite{gusynNJL} were considering a reducible representation of the gamma matrices, which lead them to an overall factor of 2 with respect to (\ref{partfunct}). But, more important, in the same reference the determinant of the Dirac operator was evaluated as the square root of minus the squared Dirac operator, thus not including the factor coming from the phase of the determinant or, equivalently, from the multiplicative anomaly \cite{kontsevich}, which we did consider (see equation (\ref{delta1})). As we will discuss in detail in section \ref{Sect6}, this leads to completely different predictions regarding the Hall conductivity.

\section{Free energy and particle density}\label{Sect5}

 From equation (\ref{partfunct}), the free energy per
 unit area ($F=-\frac{1}{\beta}\log{Z}$) can be obtained
 \footnote{Consistently with the comments in previous sections, all the results
 in this section and in the rest of this paper are independent from the representation of the gamma matrices chosen.}. It is given
by \beq \fl\nn F(\mu, B, \beta)&=& -\Delta_L
\left\{\frac{1}{\beta}\log{\left(2\cosh{\left(\frac{\mu
\beta}{2}\right)}\right)}-\frac{|\mu|}{2}\right.+
 \sqrt{2e B}\zeta_R \left(-\frac{1}{2}\right)\\\fl &+&
\left.\frac{1}{\beta}\sum_{n=1}^{\infty} \log{\left(1+
e^{-2\sqrt{2ne B}\beta}+2\cosh{(\mu \beta )}e^{-\sqrt{2ne
B}\beta}\right)}\right\},\label{Fmenor}\eeq

Moreover, the free energy is continuous at ${\mu}^2 =2k e B,
k=0,...,\infty$. In the low-temperature limit one has \beq\fl \nn
F(2ke B<{\mu}^2 < 2(k+1)e B)\rightarrow_{\beta\rightarrow
\infty}-\Delta_L \left\{\sqrt{2e
B}\left(\zeta_R
\left(-\frac{1}{2}\right)-\sum_{n=1}^{k}\sqrt{n}\right)+k|\mu|\right\}\,.\eeq

Note that, for $k=0$, this result coincides with the Casimir
energy obtained in section \ref{Sect2}, even for $\mu\neq0$, but
in this range, i.e., for $\mu$ less than the first Landau level,
if positive, or greater than minus the first Landau level, if
negative.

The mean particle density can be obtained, also from (\ref{partfunct}), as
$N=\frac{1}{\beta}\frac{d}{d\mu}\log{Z}$. For nonzero temperature
and arbitrary $\mu$ one has
\beq \nn N(\mu, B, \beta)&=&\Delta_L
\left\{\frac{1}{2}\left[\tanh{(\frac{\mu\beta}{2})}-sign(\mu)\right]\right.\\
&+&\left.\sum_{n=1}^{\infty} \frac{2\sinh{(\mu\beta)}e^{-\sqrt{2ne B}\beta}}{1+
e^{-2\sqrt{2ne B}\beta}+2\cosh{(\mu \beta
)}e^{-\sqrt{2ne B}\beta}}\right\}\,,\label{Nmenor}\eeq

For nonvanishing $\mu$, the low-temperature
limit differs, depending on the $\mu$-range considered \beq \nn
N(2ek B<{\mu}^2 < 2e (k+1)B)\rightarrow_{\beta\rightarrow
\infty}k\Delta_L\, sign(\mu)\,,\eeq where, as before,
$k=\left[\frac{{\mu}^2}{2e B}\right]$.

This result was to be expected for particles with the
statistic of fermions, since relativistic field theory naturally
leads to the spin-statistics theorem. At zero temperature, $\mu$
is nothing but the Fermi energy; for example, for $\mu>0$, as
$\mu$ grows past a Landau level, such level becomes entirely
filled.

From the previous result, the mean value of the particle density
at zero temperature can be obtained. After recovering units, one
has \beq \nn  j^0(2e c^2\hbar Bk<{\mu}^2 < 2e B
c^2\hbar(k+1))=\frac{-kce^2B}{h}\, sign(\mu)\,,\eeq the other two
components of the current density tri-vector being equal to zero
in the absence of an electric field.
Now, the zero-temperature limit of the same tri-vector in the
presence of crossed homogeneous electric ($F^{\prime}$) and
magnetic ($B^{\prime}$) fields can retrieved, for $F^{\prime}<c
B^{\prime}$, by performing a Lorentz boost with absolute value of
the velocity $\frac{F^{\prime}}{B^{\prime}}$. Suppose, for
definiteness, that the homogeneous electric field points along the
positive $y$ axis. Then, the velocity of the Lorentz boost must
point along the negative $x$-axis, and the transformation gives as
a result \beq \nn {j^{\prime}}^0=\frac{-nce^2B^{\prime}}{h}\,
sign(\mu)\,,\quad{j^{\prime}}^{x}=\frac{-ne^2F^{\prime}}{h}\,
sign(\mu)\,,\quad{j^{\prime}}^{y}=0\,.\eeq

As a consequence, the quantized zero-temperature Hall conductivity
is \beq \nn \sigma_{xy}= \frac{-ne^2}{h}\, sign(\mu)\,.\eeq

\section{Final comments}\label{Sect6}

A comment is in order concerning the zero temperature value of the particle density in equation (\ref{Nmenor}) when the chemical potential coincides with an energy level, i.e., for ${\mu}^2 = 2e kB, \, k=0,...,\infty$
In all these cases, the operation of setting $\mu$ to its value doesn't commute with that of taking the $\beta\rightarrow \infty$ limit.

For instance, in the case $\mu=0$, $N(0, B, \beta)$ is undefined at all temperatures, while $\lim_{\mu\rightarrow 0}\left[\lim_{\beta\rightarrow \infty}N({\mu}^2 < 2e B, B, \beta)\right]=0$.

On the other hand, each time ${\mu}^2=2e kB, \quad k=1,...,\infty$, one has $\lim_{\beta\rightarrow \infty} N(\pm \sqrt{2e kB}, B, \beta)=\pm \Delta_L (k-\frac12)$, while $\lim_{\mu\rightarrow\pm \sqrt{2e kB}}\left[\lim_{\beta\rightarrow \infty} N(\mu, B, \beta)\right]$ is undetermined.

\bigskip

As already commented in the Introduction, after this calculation was finished, reference \cite{gusynhall} was brought to our attention. Their prediction concerning the Hall conductivity differ from ours. In fact, the Hall conductivity in that paper contains an overall factor of $4$ with respect to ours, which is due to the use of a reducible representation and an extra sum over two values of spin. This is a more or less trivial difference. But, more important, after this factor is removed, their Hall conductivity is quantized in half-integer units of magnetic flux. This originates the word ``unconventional" in the title of \cite{gusynhall}. Our calculation, instead, leads to a Hall conductivity quantized in terms of integer quanta of flux density, which is entirely conventional, but obtained here from very first principles. The difference between both results is due to the inclusion, in our calculation, of the phase of the determinant (or, equivalently in this case, of the multiplicative anomaly). The interesting comment at this point is that experimental results on the Integer Quantum Hall Effect can clarify the physical relevance of the multiplicative anomaly (some recent results \cite{nature1,nature2} seemingly favor ``unconventional" quantization).

Finally, we mention that the more realistic case of massive
fermions is at present under study \cite{BGSS}.

\bigskip

\ack{We thank Professors Paola Giacconi, Gerardo Rossini and Roberto Soldati for useful discussions. We also thank Professor V.P. Gusynin for bringing references \cite{gusynhall} and \cite{gusynNJL} to our knowledge.\\ This work was partially supported by Universidad
Nacional de La Plata, under Grant 11/X381.}

\end{document}